\documentclass[a4paper, amsfonts, amssymb, amsmath, reprint, showkeys, nofootinbib, twoside]{revtex4-1}
\pdfoutput=1
\usepackage[english]{babel}
\usepackage[utf8]{inputenc}
\usepackage{textcomp}
\usepackage{enumitem}
\usepackage{booktabs}
\AtBeginDocument{
\heavyrulewidth=.08em
\lightrulewidth=.05em
\cmidrulewidth=.03em
\belowrulesep=.65ex
\belowbottomsep=0pt
\aboverulesep=.4ex
\abovetopsep=0pt
\cmidrulesep=\doublerulesep
\cmidrulekern=.5em
\defaultaddspace=.5em
}
\usepackage[colorinlistoftodos, color=green!40, prependcaption]{todonotes}
\usepackage[pdftex, pdftitle={Article}, pdfauthor={Author}]{hyperref} 
\begin{document}
\title{Comment on\\ ``Capturing Phase Behavior of Ternary Lipid Mixtures 
  with a Refined Martini Coarse-Grained Force Field''}

\author{Matti Javanainen}
    \email[Correspondence email address: ]{matti.javanainen@gmail.com}
    \affiliation{Institute of Organic Chemistry and Biochemistry, Czech Academy of Sciences, Flemingovo n\'{a}m. 542/2, 160 00 Prague 6, Czech Republic}
\author{Balazs Fabian}
    \affiliation{Institute of Organic Chemistry and Biochemistry, Czech Academy of Sciences, Flemingovo n\'{a}m. 542/2, 160 00 Prague 6, Czech Republic}
\author{Hector Martinez-Seara}
    \affiliation{Institute of Organic Chemistry and Biochemistry, Czech Academy of Sciences, Flemingovo n\'{a}m. 542/2, 160 00 Prague 6, Czech Republic}

\date{\today} 

\begin{abstract}
    We report here on the pitfalls of the simulation model introduced in the 
    ``Capturing Phase Behavior of Ternary Lipid Mixtures with a Refined Martini 
    Coarse-Grained Force Field'' [Journal of Chemical Theory and Computation  2018, 14, 11, 6050–6062].
    This refined Martini model was reported to reproduce experimental phase diagrams for a 
    ternary DOPC/DPPC/cholesterol mixture, including the coexistence of two liquid phases. 
    However, we demonstrate that this coexistence only emerged due to an unfortunate choice 
    of simulation parameters, which leads to poor energy conservation. Specifically,
    the constraints on the cholesterol model drained energy out from the membrane,
    resulting in two coexisting phases at drastically different temperatures. 
    Using the simulation parameters recommended for the used cholesterol model, 
    this artefact is eliminated, yet so is phase coexistence, \textit{i.e.} experimental 
    phase diagrams are no longer reproduced.
    
    \textit{It is important to highlight that the present comment was submitted to 
    Chemical Theory and Computation. However, it was rejected without peer-review
    by the Editor-in-Chief, who stated that the journal ``rarely publishes such material''.}
\end{abstract}


\maketitle

At room temperature and at certain ratios, the canonical mixture of 
dipalmitoylphosphatidylcholine (DPPC) with two saturated acyl chains, 
dioleoylphosphatidylcholine (DOPC) 
with two monounsaturated acyl chains, and cholesterol (CHOL) undergoes phase separation 
into liquid ordered (L\textsubscript{o}) and liquid disordered (L\textsubscript{d}) 
phases \cite{veatch2003separation}. It is well-documented that the standard 
implementation of the coarse-grained (CG) Martini model \cite{marrink2007martini} 
does not capture this behavior \cite{carpenter2018capturing,domanski2012transmembrane,
davis2013predictions}. Therefore, phase separation and phase coexistence studies 
performed using the Martini model replace DOPC by dilinoleoylphosphatidylcholine 
(DLiPC) with two diunsaturated acyl chains as low transition temperature ($T_\mathrm{m}$) 
lipid \cite{risselada2008molecular}. Unfortunately, no experimental phase diagrams 
for the DPPC/DLiPC/CHOL mixture exist to our knowledge, rendering the comparison 
between simulations and experiments qualitative at best.

Recently, \citeauthor{carpenter2018capturing} tackled this outstanding issue by a
careful refinement of the Martini parameters for DOPC and DPPC. They adjusted the 
bonded parameters of each bead separately to reproduce the bond length and angle
distributions extracted from atomistic simulation data \cite{carpenter2018capturing}. 
This approach diverts from the building block concept of the version 2 family of the 
Martini force field, where the number of bonded parameters is minimized. With this 
extended freedom to fine-tuning of the interactions, \citeauthor{carpenter2018capturing} 
achieved almost quantitative agreement with the experimental phase diagrams for the 
mixture of DPPC, DOPC, and CHOL \cite{veatch2003separation,davis2009phase}. Considering 
the known limitations of the CG models \cite{alessandri2019pitfalls}, this is an 
impressive milestone on the way to accurate modeling of complex lipid mixtures.
In their work, \citeauthor{carpenter2018capturing} derived two different models; 
``Extensible parameters'' where DPPC and DOPC head groups share identical interaction 
parameters, and ``Optimal parametes'', where they are allowed to differ. The latter 
parameter set is of more interest, as it provides the best agreement with the 
experimental phase diagram \cite{carpenter2018capturing}.

Here, we demonstrate that the reported liquid--liquid coexistence in the model by 
\citeauthor{carpenter2018capturing} (``Optimal parameters'') is an artifact caused 
by an unfortunate choice of simulation parameters for GROMACS. Namely, the used current 
cholesterol model by \citeauthor{melo2015parameters} \cite{melo2015parameters} was 
parametrized using a very conservative set of parameters for the LINCS constraint 
algorithm to properly model the ring structure of cholesterol.  Unfortunately, 
these more conservative constraint options were not followed by 
\citeauthor{carpenter2018capturing} \cite{carpenter2018capturing}, who instead
followed the parameter set generally recommended for the Martini model. When the 
cholesterol model is simulated with these generally used parameters, \textit{i.e.} 
4\textsuperscript{th} order expansion of the constraint coupling matrix 
(\texttt{lincs\_order} equal to 4) and 1 iteration of constraints per integration 
time step (\texttt{lincs\_iter} equal to 1), energy is not properly conserved 
due to the presence of virtual sites and corresponding coupled constraints in 
this model. This effect, which is also highlighted in theGROMACS manual 
\cite{abraham2015gromacs}, manifests itself especially at large integration time 
steps. 

In the case of the study by \citeauthor{carpenter2018capturing} 
\cite{carpenter2018capturing}, CHOL constraints drain out a substantial amount of 
energy from their neighborhood, which in the case of the studied ternary lipid mixtures 
consists preferentially of DPPC. Thus, this leads to the cooling down of CHOL and the 
surrounding DPPC lipids which together form a sort of ordered phase. In contrast --- 
to maintain the target temperature of the thermostat --- DOPC-rich region gets warmer 
than the target temperature of the thermostat, thus forming a very disorered phase. 
This Maxwell demon causes the temperatures of DPPC (and CHOL) and DOPC to diverge 
exponentially as a function of the integration time step from the target temperature 
of the thermostat, therefore resulting in the apparent L\textsubscript{o} phase being 
dozens of K cooler than the L\textsubscript{d} phase. Importantly, 
L\textsubscript{o}/L\textsubscript{d} coexistence regions reported by the authors 
and matching the experimental phase diagram in the DPPC/DOPC/CHOL mixtures are only 
recovered in simulations in the presence of this artifact.

We show that this build-up of temperature difference can be at least partially 
prevented by either 1) following the suggested LINCS parameters for CHOL simulations 
by \citeauthor{melo2015parameters}, 2) coupling the different lipid types to separate 
thermostats, or 3) decreasing the simulation time step. All these options, that 
improve energy conservation, cause the L\textsubscript{o}/L\textsubscript{d} phase 
coexistence to vanish. Instead, an almost ideal mixing is instead observed.

Thus, when achieving a proper NPT ensemble, the model by 
\citeauthor{carpenter2018capturing} does not present an improvement over the 
standard Martini v2.2 lipid model in capturing the phase coexistence in DOPC/DPPC/CHOL 
mixtures \cite{davis2013predictions,risselada2008molecular,carpenter2018capturing}.
Therefore, further work is required before a direct comparison of experimental and 
Martini-based coarse grained simulations on phase-separating lipid mixtures is viable. 
Meanwhile, the DPPC/DLiPC/CHOL mixture in the standard Martini implementation 
will---despite its obvious limitations---serve as the main model of choice for 
studies on L\textsubscript{o}/L\textsubscript{d} phase coexistence.

\section{Results}

\begin{figure*}[htb]
    \begin{center}
    \includegraphics[width=0.85\textwidth]{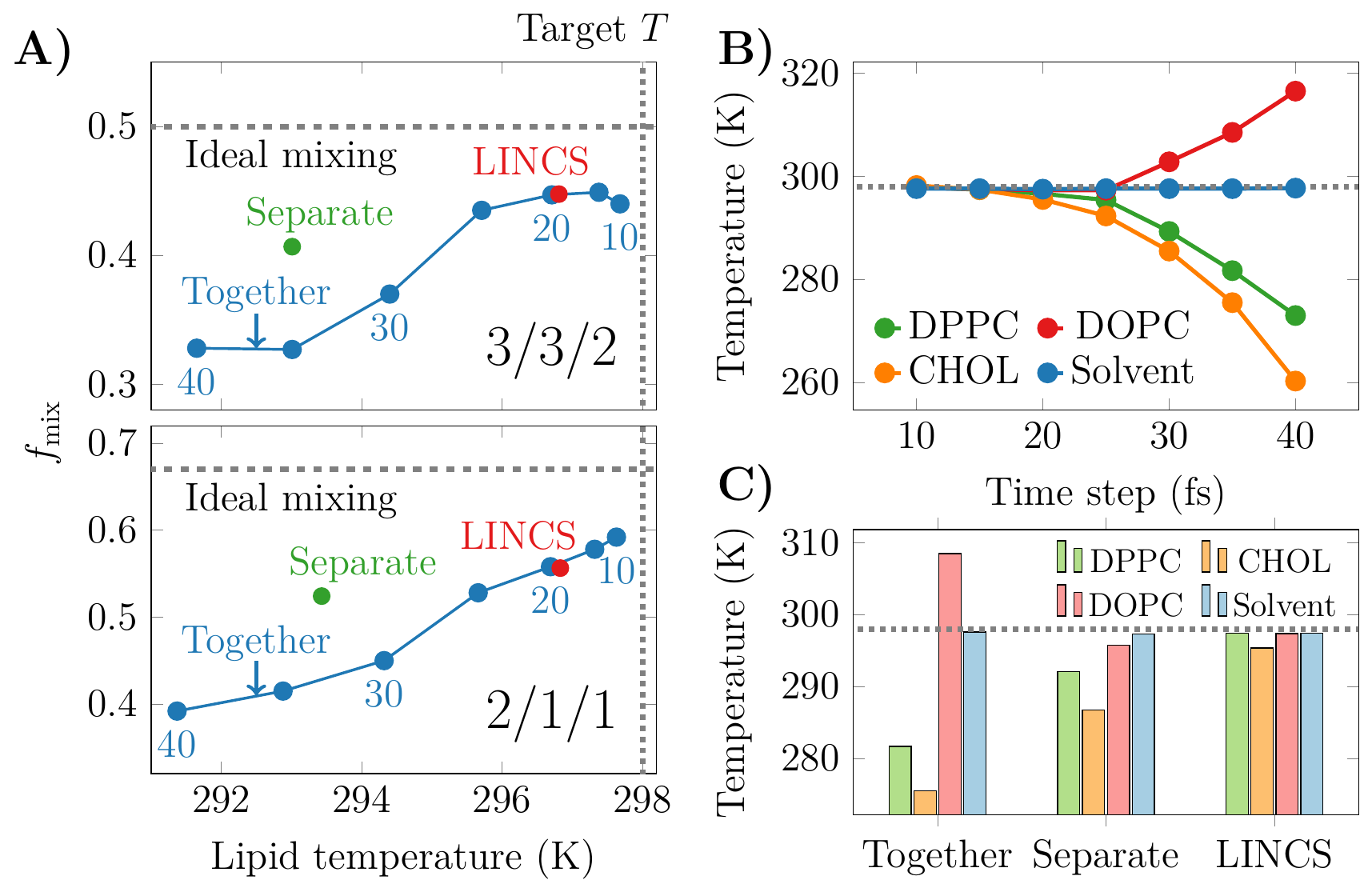}
    \caption{\label{fig:fig1}%
    LINCS settings and temperature coupling affect the degree of phase separation. 
    A) The degree of phase separation as a function of the simulation temperature of 
    the lipids, both of which depend on the integration time step, indicated by 
    point labels. Only the labels that are multiples of 10 are shown for clarity. 
    The dotted gray line shows the target temperature of the 
    thermostat. Data for the two studied mixtures are shown in the two panels,
    and the $f_\mathrm{mix}$ values corresponding to ideal mixing are highlighted
    by dashed gray lines. The blue markers show data for simulations in which all
    lipids are coupled \textbf{Together} to the thermostat (Set 1 in 
    Table~\ref{tab:simulations}). Green markers show data for simulations in which
    lipid types are coupled to \textbf{Separate} thermostats (Set 2 in 
    Table~\ref{tab:simulations}). Red markers show data for simulations in which
    more conservative \textbf{LINCS} parameters are used (Set 3 in 
    Table~\ref{tab:simulations}).
    B) Temperatures of the lipid types and the solvent beads as a function
    of the integration time step. Data are only shown for the 3/3/2 mixture, yet
    the behavior in the 2/1/1 mixture is essentially identical.
    C) Temperatures of the lipid types and the solvent beads as a function of 
    temperature coupling and LINCS options. Data are again shown for the 3/3/2
    mixture only.
    }
    \end{center}
\end{figure*}

To demonstrate the issues in the model by \citeauthor{carpenter2018capturing}, 
we first followed the original simulation protocol reported in  
Ref.~\citenum{carpenter2018capturing}. We performed 15~\textmu{}s simulations of 
3/3/2 and 2/1/1 compositions of the DPPC/DOPC/CHOL mixture (see Methods and Set 
1 in Table~\ref{tab:simulations}). Both of these compositions fall in the heart 
of the L\textsubscript{o}/L\textsubscript{d} coexistence region 
\cite{carpenter2018capturing}. The simulations were performed at 298~K using time 
steps of 10, 15, 20, 25, 30, 35, and 40~fs. 
Importantly, 35~fs was used in the original work \cite{carpenter2018capturing}. 
We characterized the degree of phase separation using the mean contact fraction 
\cite{domanski2012transmembrane} ($f_\mathrm{mix}$, see Methods) extracted from 
the last 1~\textmu{}s simulation. For each value of the integration time step, 
we extracted $f_\mathrm{mix}$  and the temperature of the lipids. With 
blue markers in Fig.~\ref{fig:fig1}A), 
we demonstrate the obvious correlation of $f_\mathrm{mix}$ of the function of lipid 
temperature, both of which depend on the used integration time step. Notably, larger 
time steps lead to significantly cooler membranes ($<$ 298~K) and smaller values 
of $f_\mathrm{mix}$, \textit{i.e.} higher degree of phase separation. With time 
steps that maintain the overall temperature of the lipids reasonably close to the 
target temperature of 298~K, both mixtures display nearly ideal mixing instead. 
The fact that the target temperature of the thermostat is not preserved even with 
small time steps is in line with a recent report on the imperfect energy conservation 
with the New-RF simulation settings used \cite{benedetti2018comment}.

We also extracted the temperatures of the different lipid types. As this 
information not available in the original trajectories as no velocities were saved, 
we extended each simulation by 100~ns, during which we also saved the instantaneous 
velocities (see Methods). The temperatures of each lipid type are shown as a function 
of the integration time step in Fig.~\ref{fig:fig1}B). It is evident that the increase 
in time step causes the low-$T_\mathrm{m}$ lipid (DOPC) to heat up and the 
high-$T_\mathrm{m}$ (DPPC) lipid to cool down. The behaviors of DPPC and CHOL are 
similar. For time steps up to 20~fs, this effect is negligible, yet with the 35~fs 
time step used by \citeauthor{carpenter2018capturing}, the temperature difference 
between the lipid components is already more than 20~K with DPPC being $\sim$30~K 
below its $T_\mathrm{m}$ of 314~K. Thus, it is clear that decreasing the integration 
time step improves energy conservation, yet leads to a smaller level of phase separation. 

To pinpoint the culprit of the poor energy conservation, we performed additional 
simulations with the input parameters slightly modified. First, we used the same 
integration time step of 35~fs as \citeauthor{carpenter2018capturing} but with the 
LINCS parameters provided by \citeauthor{melo2015parameters} 
\cite{melo2015parameters}---namely two LINCS iterations per time step, and an 
8\textsuperscript{th} order expansion of the constraint coupling matrix (Set 3 in 
Table~\ref{tab:simulations}). Data for these settings is shown in red in 
Fig.~\ref{fig:fig1}A). It is evident that this more accurate handling of the 
constraints leads to a much smaller degree of cooling of the lipids. With these 
settings, phase separation also essentially vanishes. The temperatures of each 
lipid type in the simulations with a time step of 35~fs and with various simulation 
settings are shown in Fig.~\ref{fig:fig1}C). The more accurate LINCS settings clearly 
lead to all the lipids being at almost the same temperature which also closely matches 
the target one. This indicates that energy conservation can be improved by at least 
these two ways---decreasing the time step or increasing the number of LINCS iterations 
and the order of the LINCS expansion. The fact that ``LINCS'' and ``20'' overlap in 
Fig.~\ref{fig:fig1}A) suggests that the effect on phase separation is independent on 
the means of improving energy conservation. However, both approaches lead to a 
disagreement between the simulation and the experimental phase diagram, contrary 
to the results reported in the paper by \citeauthor{carpenter2018capturing}.

We performed an additional simulation, in which we attempted to improve energy 
conservation by using a separate thermostat for each of the lipid types while still 
using an integration time step of 35~fs (Set 2 in Table~\ref{tab:simulations}). As 
shown in Fig.~\ref{fig:fig1}A) by green dots, these settings lead to an almost as 
cool membrane as in the case when all lipids are coupled to the same thermostat. 
Still, the degree of phase separation is significantly
decreased. Based on Fig.~\ref{fig:fig1}C), the used temperature coupling does
not prevent CHOL and DPPC from cooling down, yet it does not allow DOPC to
heat up. Therefore, the average lipid temperature decreases, whereas the 
temperature difference between DPPC and DOPC is minor resulting in no phase 
separation.

\begin{figure*}[htb!]
    \begin{center}
    \includegraphics[width=0.85\textwidth]{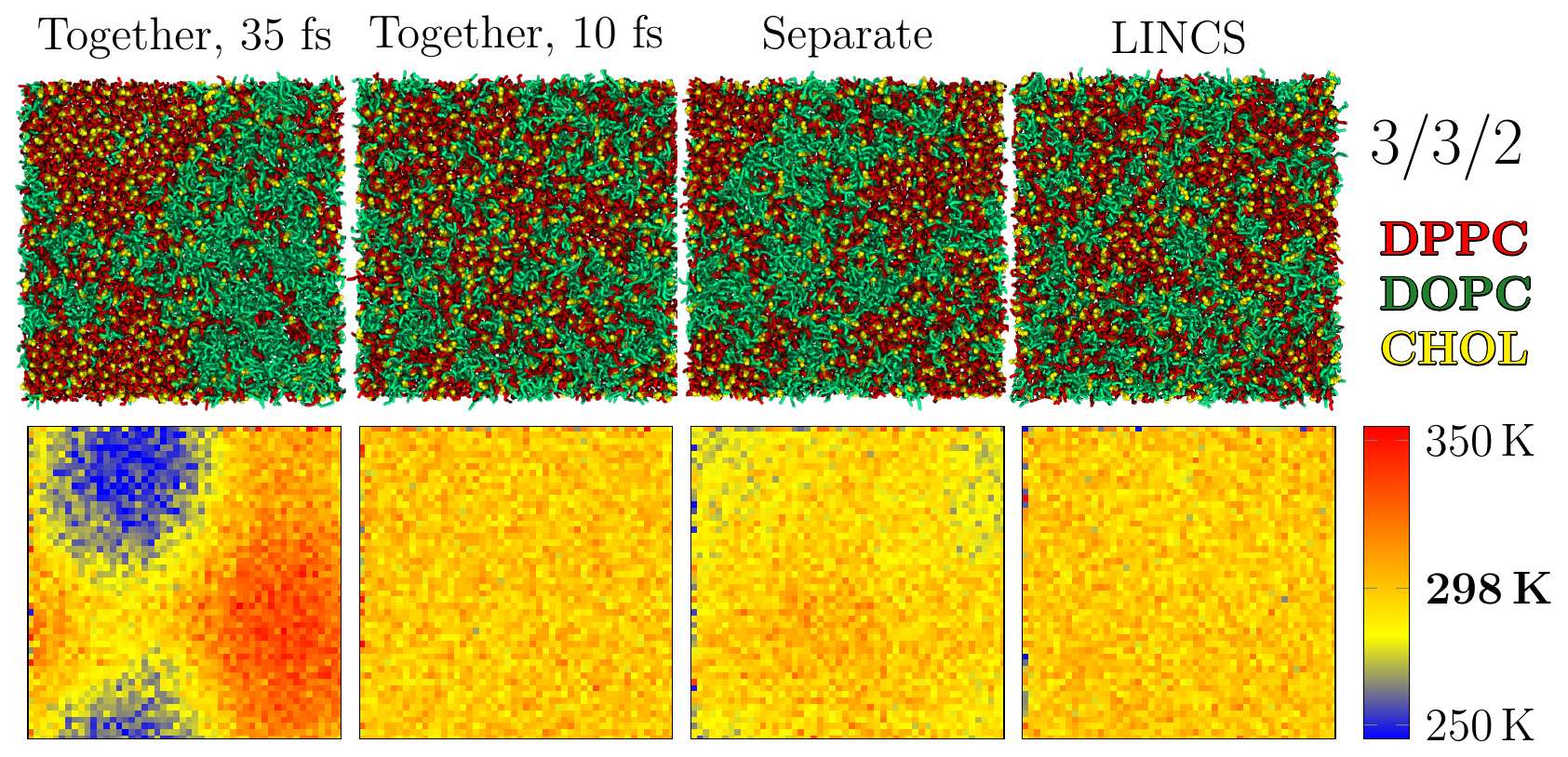}
    \caption{\label{fig:fig2}%
    Top row: final structures of chosen simulations with the 3/3/2 
    DPPC/DOPC/CHOL mixture after 15~\textmu{}s. DPPC, DOPC, and CHOL
    are shown in red, green, and yellow, respectively. Only the system
    where all lipids are coupled to the same thermostat (``Together'')
    and a 35~fs integration time step is used undergoes phase separation.
    Some heterogeneity is also seen in other systems with DPPC and CHOL
    clustering together. 
    Bottom row: the temperature maps calculated from the 100~ns simulations
    during which velocities are saved (see Methods). For the 
    ``Together, 35~fs'' system, the two phases show almost 100~K 
    difference in temperature. When the lipid types are coupled to separate 
    thermostats (``Separate''), the DPPC-rich regions are slightly cooler 
    than the remainder of the membrane. For the other systems, the 
    temperature is close to the target on (298~K) in the entire membrane.
    }
    \end{center}    
\end{figure*}

We also plot the final structures of selected simulations on the top row of 
Fig.~\ref{fig:fig2}. It is evident that proper phase separation takes place
with a time step of 35~fs and following the simulation options of 
\citeauthor{carpenter2018capturing}\cite{carpenter2018capturing} (Set 1 in
Table~\ref{tab:simulations}), whereas decreasing the time step to 10~fs 
results in an almost ideal mixing of the lipid types. Similar behavior is 
observed with the conservative LINCS parameters (Set 3 in 
Table~\ref{tab:simulations}). However, for the case in which all lipid
types are coupled separately to a thermostat (Set 2 in 
Table~\ref{tab:simulations}), some heterogeneity is still observed. The local 
temperature maps, calculated from the 100~ns simulation during which velocities 
were saved, are shown on the bottom row of Fig.~\ref{fig:fig2}. It is evident 
that the observed phase separation with the model by 
\citeauthor{carpenter2018capturing} goes hand in hand with the uneven temperature 
distribution in the membrane, indicating that the systems is not sampling the 
proper equilibrium ensemble. With the simulation setup of 
\citeauthor{carpenter2018capturing} \cite{carpenter2018capturing}, the 
temperatures of the L\textsubscript{o} and L\textsubscript{d} phases can differ 
as much as 100~K between regions.

\section{Conclusions}

From the data presented above, we can draw the following conclusions: 
\textbf{1)} The use of LINCS parameters that are incompatible with the used 
CHOL model, combined with a large integration time step, results in poor energy 
conservation. In the case of DPPC/DOPC/CHOL mixture considered by
\citeauthor{carpenter2018capturing} \cite{carpenter2018capturing}, the virtual 
site model of CHOL \cite{melo2015parameters} drains out a significant amount of 
energy from the system, mainly from DPPC which locates itself close to CHOL.
\textbf{2)} When all lipids are coupled together to the thermostat, as was done 
by \citeauthor{carpenter2018capturing}\cite{carpenter2018capturing}, the cooling of 
CHOL and DPPC leads to the heating up of DOPC. This temperature difference
drives the separation of the lipids into hot (DOPC-rich) and cool (DPPC- and 
CHOL-rich) phases, whose temperatures differ by significantly. This behavior 
is in obvious violation of the second law of thermodynamics.
\textbf{3)} The poor energy conservation can be cured by either decreasing the
integration time step or using the LINCS parameters that were used in the
parametrization of the virtual site model of CHOL \cite{melo2015parameters}.
This better energy conservations leads to the loss of phase coexistence, and
thus to poor agreement with experimental phase diagrams. Additionally, by coupling 
all lipid types to separate thermostats, the temperature difference between DPPC and 
DOPC is limited, and only a small degree of lipid demixing is observed.

It is also worth mentioning that the standard implementation of the current version
2.2 of the Martini model does not display phase separation of the DPPC/DOPC/CHOL
mixture even when the simulation settings of \citeauthor{carpenter2018capturing} 
are used (set 4 in Table~\ref{tab:simulations}). This results from the fact that
the standard Martini DOPC model mixes fairly well with cholesterol. Indeed, 
a hybrid DPPC/DOPC/CHOL mixture with the parameters of 
\citeauthor{carpenter2018capturing} used for DPPC and the standard Martini
parameters used for DOPC did not phase-separate. However, with DOPC parameters
adapted from \citeauthor{carpenter2018capturing} and DPPC parameters from the
standard Martini implementation, phase separation was recovered. This indicates
that the unfavorable mixing of DOPC and CHOL in the model by 
\citeauthor{carpenter2018capturing} is central for the co-cooling of DPPC
with CHOL, which is required for the formation of a cool L\textsubscript{o}
phase.

All in all, while the model of \citeauthor{carpenter2018capturing} seemed to finally 
reproduce the experimental phase diagram  of the well-studied DPPC/DOPC/CHOL lipid 
mixture, our results demonstrate that it does due to a simulation artifact. This 
artifact relates to a poor energy conservation, which results from an unfortunate 
combination of large integration time steps and not using the the LINCS parameters 
required by the CHOL model. These choices lead to a clear violation of target NPT 
ensemble. Unfortunately, this compromises the validity of any findings obtained with
the model by \citeauthor{carpenter2018capturing} \cite{carpenter2018capturing}, and
suggests that further work is still needed until a Martini-like coarse-grained model 
reproduces the phase behavior of the canonical DPPC/DOPC/CHOL mixture.

\section{Methods}

\subsection{Simulations}

In the first set of simulations, we strictly followed the protocol of 
\citeauthor{carpenter2018capturing} \cite{carpenter2018capturing} and built
lipid membranes with dimensions of $30\times30\times15$~nm$^2$, which had
a total of 3040 lipids and $\sim$77000 solvent beads, out of which 10\% were 
modeled as antifreeze particles. The systems were set up by the \texttt{insane} 
tool \cite{wassenaar2015computational}. We considered two compositions in 
the heart of the L\textsubscript{o}/L\textsubscript{d} coexistence region, 
namely DPPC/DOPC/CHOL ratios of 3/3/2 and 2/1/1. We applied restraints with a 
force constant of 2~kJ/(mol$\times$nm$^2$) to the phosphate beads of the 
phospholipids in one leaflet. All in all, we used the same equilibration protocol 
and simulation parameters as \citeauthor{carpenter2018capturing} in the 
commented paper \cite{carpenter2018capturing}. Namely, the New-RF simulation 
parameters \cite{de2016martini} were used, unless otherwise mentioned. To 
pinpoint the violation of the NPT ensemble to the used CHOL model, we performed 
the following simulations (see also Table~\ref{tab:simulations}):

First, following \citeauthor{carpenter2018capturing}, the lipids and the 
solvent were separately coupled to a thermostat with a target temperature 
of 298~K, the 15~\textmu{}s simulations with time steps of 10, 15, 20, 25, 30, 
35, and 40~fs, saving the trajectories every 1~ns. Secondly, we performed 
15~\textmu{}s simulations with a 35~fs time step but with all the lipid 
types (DPPC, DOPC, and CHOL) coupled separately to a thermostat. Thirdly,
we performed 15~\textmu{}s simulations with two iterations of the LINCS 
algorithm (\texttt{lincs\_iter = 2}) and using the 8\textsuperscript{th} 
order expansion of the constraint coupling matrix 
(\texttt{lincs\_order = 8}). Here, all lipids were coupled together to 
the thermostat.

We also run additional 
100~ns simulations, starting from the final structures of the aforementioned 
simulations, and saved the velocities (.trr file) so that the temperatures 
of each lipid type could be extracted using \texttt{gmx traj}. As the used
cholesterol model has constraints, we corrected for the missing degrees
of freedom. 

\begin{table}[htb]
    \begin{center}
    \caption{\label{tab:simulations}%
        Simulated systems. Composition is given as DPPC/DOPC/CHOL. 
        ``$\Delta t$'' is the integration time step in fs. ``$T$-coupl.'' refers to
        the way the lipid temperatures are coupled; the lipids are being coupled either
        together to one thermostat or separately to three thermostats. 
    }
    \begin{tabular}{ccccc}
        \toprule
        Composition & $\Delta t$ & $T$-coupl. & \texttt{lincs\_order} & \texttt{lincs\_iter} \\
        \midrule
        \multicolumn{5}{l}{Set 1. Lipids coupled \textbf{Together}}\\
        \midrule
        3/3/2 & 10 & Together & 4 & 1 \\
        3/3/2 & 15 & Together & 4 & 1 \\
        3/3/2 & 20 & Together & 4 & 1 \\
        3/3/2 & 25 & Together & 4 & 1 \\
        3/3/2 & 30 & Together & 4 & 1 \\
        3/3/2 & 35 & Together & 4 & 1 \\
        3/3/2 & 40 & Together & 4 & 1 \\
        \midrule
        2/1/1 & 10 & Together & 4 & 1 \\
        2/1/1 & 15 & Together & 4 & 1 \\
        2/1/1 & 20 & Together & 4 & 1 \\
        2/1/1 & 25 & Together & 4 & 1 \\
        2/1/1 & 30 & Together & 4 & 1 \\
        2/1/1 & 35 & Together & 4 & 1 \\
        2/1/1 & 40 & Together & 4 & 1 \\        
        \midrule
        \multicolumn{5}{l}{Set 2. Lipids coupled \textbf{Separately}}\\
        \midrule
        3/2/2 & 35 & Separately & 4 & 1 \\        
        2/1/1 & 35 & Separately & 4 & 1 \\                
        \midrule
        \multicolumn{5}{l}{Set 3. Conservative \textbf{LINCS} settings}\\        
        \midrule
        3/2/2 & 35 & Together & 8 & 2 \\        
        2/1/1 & 35 & Together & 8 & 2 \\                
        \midrule
        \multicolumn{5}{l}{Set 4. \textbf{Unmodified} Martini v2.2}\\        
        \midrule
        3/2/2 & 35 & Together & 4 & 1 \\        
        \bottomrule
    \end{tabular}
    \end{center}    
\end{table}

\subsection{Analyses}

The contact fraction $f_\mathrm{mix}$, describing the level of lipid 
demixing, was adapted from Ref.~\citenum{domanski2012transmembrane}, and is 
defined as
\begin{align}
    f_\mathrm{mix}=\frac{c_\mathrm{US-S}}{c_\mathrm{US-S}+c_\mathrm{US-US}},
\end{align}
where for example $c_\mathrm{US-S}$ refers to contacts between unsaturated 
(US, here DOPC) and saturated (S, here DPPC) lipids. Thus, the smaller the
value of $f_\mathrm{mix}$, the sharper the separation is. A contact is 
registered if the phosphate beads of the lipids were within 1.1~nm.

The temperatures of the lipids were extracted using \texttt{gmx energy}. 
When the lipid types were coupled separately, their temperatures were 
extracted separately and the averaging was performed over the degrees
of freedom. For CHOL, this was less than $3N$ due to the constraints.
The temperatures of each lipid type were extracted from the 100~ns trajectories 
containing velocities (see Simulations Methods above) using \texttt{gmx traj}. 
The values were corrected for the missing degrees of freedom of CHOL.

The heat maps of temperature distribution were calculated by
projecting the lipid center of mass onto the macroscopic plane
of the membrane. While performing the binning, each point was 
weighted by the instantaneous temperature of the corresponding molecule.
For DPPC and DOPC, the number of degrees of freedom in each
lipid was taken as $3N$, while in the case of CHOL the decrease
of degrees of freedom in the presence of constraints was accounted for.

\begin{acknowledgments}

M.J., H.M.-S., and B.F. acknowledge support from the Czech Science Foundation 
(EXPRO grant 19-26854X). M.J. thanks the Emil Aaltonen Foundation for funding. 

\end{acknowledgments}

\newpage

\bibliography{refs}

\section{Data Availability}

The inputs and outputs for our simulations with the model by
\citeauthor{carpenter2018capturing} are available as follows:
\begin{enumerate}
    \item Long simulations with different time steps:
    \begin{itemize}[leftmargin=*]
        \item DOI: 10.5281/zenodo.3956709 (3/3/2 mixture) 
        \item DOI: 10.5281/zenodo.3956797 (2/1/1 mixture)
    \end{itemize}
    \item Short simulations with different time steps and with velocities saved:
    \begin{itemize}[leftmargin=*]
        \item DOI: 10.5281/zenodo.3956761 (3/3/2 mixture) 
        \item DOI: 10.5281/zenodo.3956812 (2/1/1 mixture)
    \end{itemize}
    \item Additional simulations of the 3/3/2 mixture with different temperature 
coupling or LINCS settings:
    \begin{itemize}[leftmargin=*]
        \item DOI: 10.5281/zenodo.3956775 (3/3/2 mixture) 
        \item DOI: 10.5281/zenodo.3956814 (2/1/1 mixture)
    \end{itemize}
\end{enumerate}

\end{document}